# Structure retrieval from 4D-STEM: statistical analysis of potential pitfalls in high-dimensional data

Xin Li[1], Ondrej Dyck[1], Stephen Jesse[1], Andrew R. Lupini[1], Sergei V. Kalinin[1*], and Mark P. Oxley[1*]

[1]Center for Nanophase Material Sciences, Oak Ridge National Laborotary

**Abstract**

Four-dimensional scanning transmission electron microscopy (4D-STEM) is one of the most rapidly growing modes of electron microscopy imaging. The advent of fast pixelated cameras and the associated data infrastructure have greatly accelerated this process. Yet conversion of the 4D datasets into physically meaningful structure images in real-space remains an open issue. In this work, we demonstrate that, it is possible to systematically create filters that will affect the apparent resolution or even qualitative features of the real-space structure image, reconstructing artificially generated patterns. As initial efforts, we explore statistical model selection algorithms, aiming for robustness and reliability of estimated filters. This statistical model selection analysis demonstrates the need for regularization and cross-validation of inversion methods to robustly recover structure from high-dimensional diffraction datasets.

# Introduction

The development of fast pixelated cameras in scanning transmission electron microscopy (STEM) now enables recording the entire scattering distribution, generating a four-dimensional (4D) dataset that consists of two real-space dimensions and two reciprocal-space dimensions for every probe position[1]–[3]. Among many other directions, much attention has been drawn to recovering real-space structure from the 4D diffraction datasets recorded by electron detectors. After the early proposals to process intensities recorded on different detector geometries by Rose [4] and Dekkers and De Lang [5], annular brightfield (ABF) [6]–[8] and differential phase contrast (DPC) [9]–[11] have become popular techniques for resolving light element atoms and atomic electromagnetic fields.

As Rose [12] pointed out, the key for obtaining a strong signal is to extract constructive inferences and suppress random noise in the convergent beam electron diffraction (CBED) pattern by segmenting detectors. For example, Findlay et al. [13] proposed an enhanced ABF imaging mode (eABF) that subtracts intensities recorded on the central circular segment in the bright field from the ABF intensity to improve contrast at light atoms. DPC for visualizing electric fields has also been recently demonstrated by segmenting detectors [14]. The introduction of pixelated detectors [1]–[3] provides great flexibility for exploring detector configurations. Pennycook et al. [1] have shown reconstruction from pixelated detectors gives rise to extremely efficient phase contrast imaging at atomic resolution, and can be used to generate contrast even with zero aberrations in STEM. Ptychographic reconstruction methods [15] and segmented detectors have also been demonstrated to improve the contrast transfer [16], [17]. The first moment (center-of-mass) of the diffraction intensities is related to the electromagnetic fields [18], [19], which can also be estimated by weighted subtraction of the intensities on multi-segment detectors [20], [21]. Recently, Seki et al. [22] proposed a theoretical framework of statistical noise for several STEM coherent imaging techniques including ABF, DPC and ptychography.

Along these lines, STEM image formation can be treated as forming a linear combination of intensities recorded by the pixelated detector. The analysis is similar for pixelated detectors and segmented detectors. To configure a DPC segmentation is a matter of estimating a filter weight for a certain region of the bright field disk to integrate. In this paper, we explored statistical model selection algorithms to find filters of CBED pixel intensities for obtaining real-space images. We illustrate that it is possible to systematically create filters that will recreate patterns in artificially prescribed images, for example, images with excessively high resolution. Thus, it is vital to cross-validate robustness and prediction performances of the estimated filters. We show that by careful choice of model selection and validation, we could get robust filters to yield comparable images to the usual high-angle angular dark field (HAADF) imaging technique.

# Formalism

Experimental 4D imaging was acquired using a Nion UltraSTEM 100 microscope operated at 60 kV accelerating voltage with a nominal 30 mrad convergence angle and an approximately 60 pA probe current. The microscope is equipped with an optically coupled Hamamatsu Orca CMOS camera with 2k by 2k pixels. In order to provide a better signal to noise ratio, the camera was binned to 256 by 256 pixels and a pixel dwell time of 20 ms was used. Preparations of CVD-grown graphene samples and in-situ dopant atom insertion follow [23], [24].

We denote the real-space image of size $I$ by $J$ as $A(i,j), i = 1, ..., I; j = 1, ..., J$ where $(i,j)$ is the scanned pixel location. We denote the associated 4D diffraction datasets as $D(i,j,k,l), k = 1, ..., K; l = 1, ..., L$ where $D(i,j,\cdot,\cdot)$ of size K by L, represents the CBED data recorded on the detector at the scanned pixel location $(i, j)$. We assume the existence of a filter $\boldsymbol{F}$ satisfying the following linear relationship:

$$A(i, j) = \sum_{k=1}^{K} \sum_{l=1}^{L} D(i, j, k, l) \times F(k, l). \quad (1)$$

We use the column vector $\boldsymbol{y} = [y_1, y_2, ..., y_n]^T, n = I \times J$ to denote the vectorized version of $\boldsymbol{A}$. We denote corresponding CBED datasets in matrix form as $\boldsymbol{X} = [\boldsymbol{x}_1, \boldsymbol{x}_2, ..., \boldsymbol{x}_n]^T$ of size $n$ by $p$, where $p = K \times L$ and column vector $\boldsymbol{x}_i$ represents a vectorized CBED image. We use the column vector $\boldsymbol{w} = [w_1, w_2, ..., w_p]^T$ to denote the vectorized filter $\boldsymbol{F}$ to be estimated. We can then rewrite the above equation in matrix form:

$$\boldsymbol{y} = \boldsymbol{X}\boldsymbol{w} \quad (2)$$

Ideally, one would like to use as few as possible training data to obtain a robust filter capable of producing high-resolution atomic-structure imaging for a broad set of CBED datasets collected under equivalent experimental conditions. Figure 1 illustrates the training datasets. Figure 1a is an experimental high angle angular dark filed (HAADF) image of size 64 by 64 probe positions, showing a graphene lattice with a 4-fold coordinated silicon atom (brighter spot). Figure 1b is an intensity image constructed with an artificially high resolution, used for training. The size of the training intensity image (Figure 1b) is 17 by 64, corresponding to the estimated atom sites in the marked area of the HAADF image (Figure 1a). Figure 1c is an example of a CBED image of size 256 by 256. Correspondingly, for training purpose, there are 17*64 CBED images associated with the marked area of the HAADF image in Figure 1a.

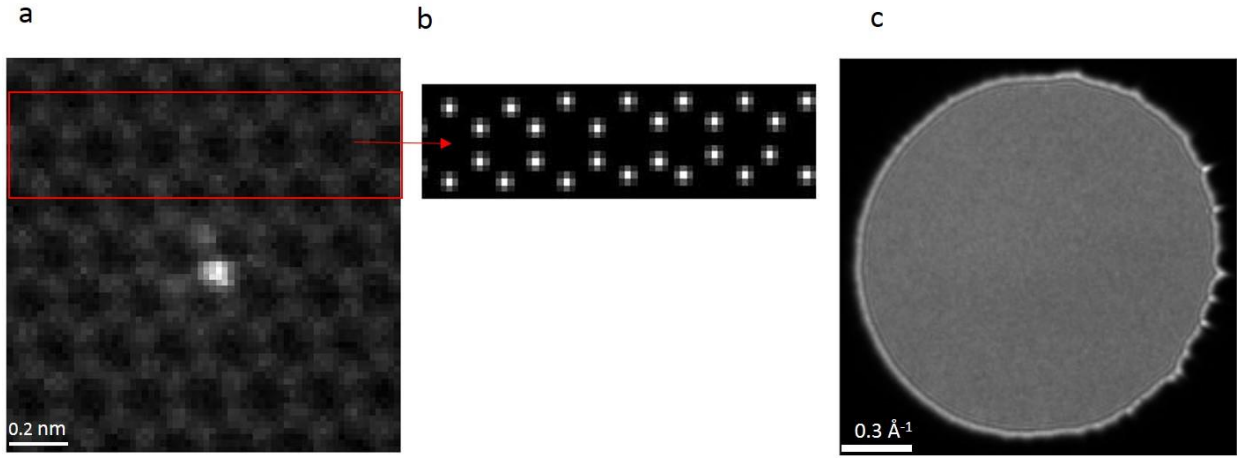

Figure 1: (a) Graphene HAADF image showing a 4-fold coordinated silicon dopant. (b) A simulated image with artificially high resolution, containing estimated atom sites in the marked area of (a). (c) An as-acquired CBED image.

Having constructed this example, we are now in a position to question the existence of such a filter $w$ that could reconstruct Figure 1b. Given the highly redundant information in CBED, it turns out we could easily find a filter (for example, via ordinary least square estimates) to reproduce the training intensity image of almost arbitrary patterns, but, with the risks of overfitting and degenerating prediction performances. Figure 2 presents some deliberately overfitted results that lead to very poor predictions, yielding meaningless interpretations. Figure 2 panels (a-d) are four (32x64 pixels) training intensity images with artificial patterns: excessively high resolution; random atom sites; rings; and a word 'caveat'. Panels (e-h) are the corresponding filters estimated from the half of the experimental CBED data (associated with upper half of the HAADF image) and the desired training images. Panels (i-l) are the images reconstructed by the filters from all (64*64) CBED patterns associated with the full HAADF image, where we can see large inconsistencies between the training (upper half) and prediction (lower half) parts.

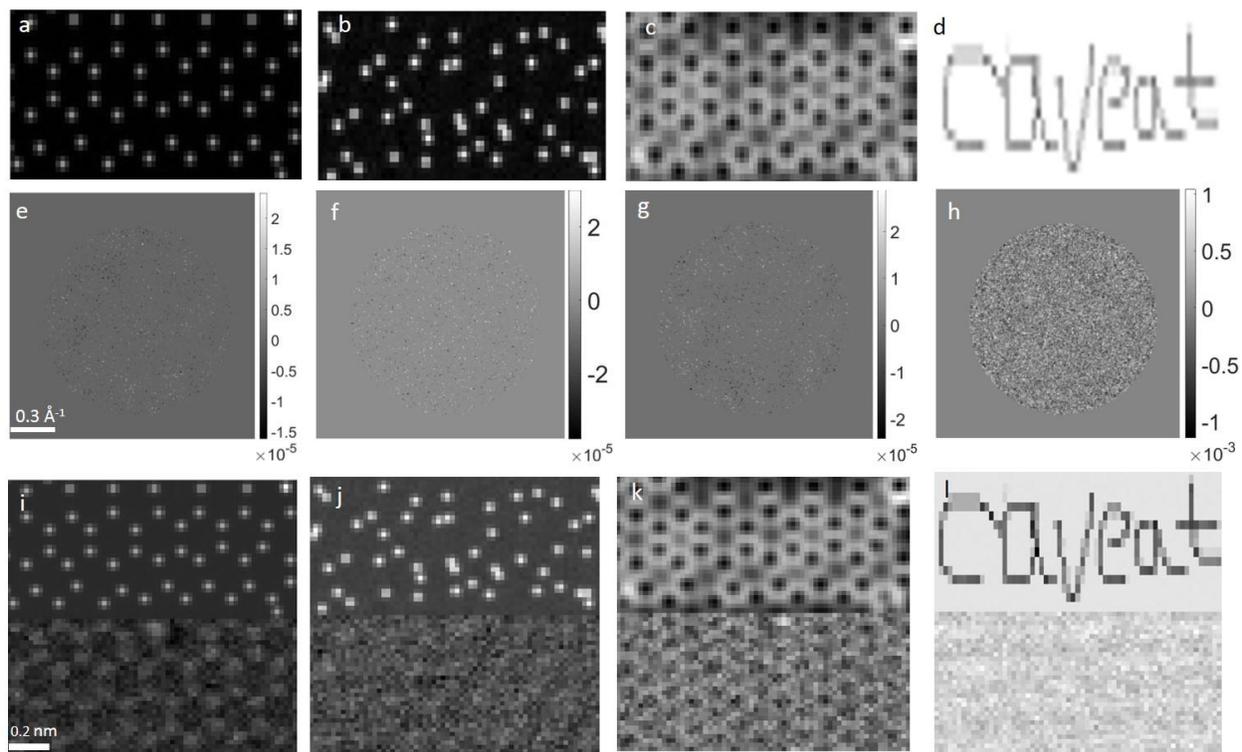

Figure 2: (a-d) Training intensity images. (e-h) Estimated filters. (i-l) Reconstructed intensity images based on all CBED patterns with the estimated filters in the second row.

In other words, it is possible to deliberately construct a filter that makes the experimental result resemble almost any desired output. Clearly there needs to be a route to evaluate the derived filters, especially for robustness and prediction performances. In fact, directly solving equation (2) is often an ill-posed problem for finding the "optimal" $w$, especially for high-dimensional variables. (In the Figure 2 examples, the variable (CBED) dimensionality $p$ = 256 X256 = 65536). Statistical model selection is a set of well-established algorithms aiming for improved performances of prediction/validation as well as model parsimony for the sake of scientific interpretations. Mathematically, the model estimator of $w$ consists of a data fidelity term and a sparsity penalty term:

$$\widehat{w} = \arg\min_w \frac{1}{2n}\|y - Xw\|^2 + \lambda P_r(w) \quad (3)$$

where

$$P_r(w) = r\|w\|_1 + \frac{(1-r)}{2}\|w\|_2^2 \quad (4)$$

here $P_r(w)$ corresponds to the elastic net sparsity regularization [25]. The elastic net model offers good flexibility on sparsity and smoothness of the solution path for $w$. Equation (3) becomes the well-known Lasso problem (L1 norm penalty) [26] when $r$ = 1, favoring sparsity in the filter $w$, meanwhile decreasing $r$ accounts for the Ridge regression (L2 norm penalty) in favor of smoothness in the filter. For a segmented electron detector with $M$ segments, the covariate matrix $X$ in equation (3) reduces to $M$ columns where the $m$th column corresponds to the integration of CBED intensities inside the $m$-th segment and the filter consists of $m$ unknown weights to be estimated.

Equation (3) can be efficiently solved by the cyclical coordinate descent algorithm [27], [28]. The coordinate-wise update computes the simple least-squares coefficients on the partial residual followed by soft-thresholding accounting for the Lasso criteria and proportional shrinkage for the Ridge penalty:

$$\widehat{w_j} \leftarrow \frac{S_{\lambda r}(\frac{1}{n}\sum_{i=1}^n X_{ij}(y_i - \widehat{y}_i^{(j)}))}{1 + \lambda(1-r)} \quad (5)$$

where

$$\widehat{y}_i^{(j)} = \sum_{l \neq j} X_{il}\widehat{w_l} \quad (6)$$

and the soft-thresholding operator $S_\gamma(z)$ is defined as:

$$S_\gamma(z) = \max\left(0, 1 - \frac{\gamma}{|z|}\right)z \quad (7)$$

In practice, the cyclical coordinate descent algorithm sequentially generates the $w$ solution path accounting for different levels of sparsity corresponding to a decreasing sequence of $\lambda$ values: each solution is used as a warm start for the next problem. Specifically, the coordinate-wise descent procedure starts from an implicitly large value $\lambda_{\max}$ for which the entire vector $\widehat{w} = 0$.

# Demonstrative Results

We first use only the information inside the bright field (BF) disk. The training dataset consists of the simulated intensity image with artificially high-resolution (Figure 1b) and the associated CBED images for marked area of the HAADF image in Figure 1a. We have examined the effect of varying r but it does not alter conclusions presented here. Without loss of generality, we set r = $5e^{-5}$ throughout the paper. We focused on the evolution of prediction performances and the corresponding geometries of estimated filters for the decreasing sequence of $\lambda$ values. Figure 3 illustrates how the sparsity of estimated filters decreases as $\lambda$ decreases, where the filling ratio is defined as the number of non-zero elements in $\widehat{w}$ over the total number of filter elements.

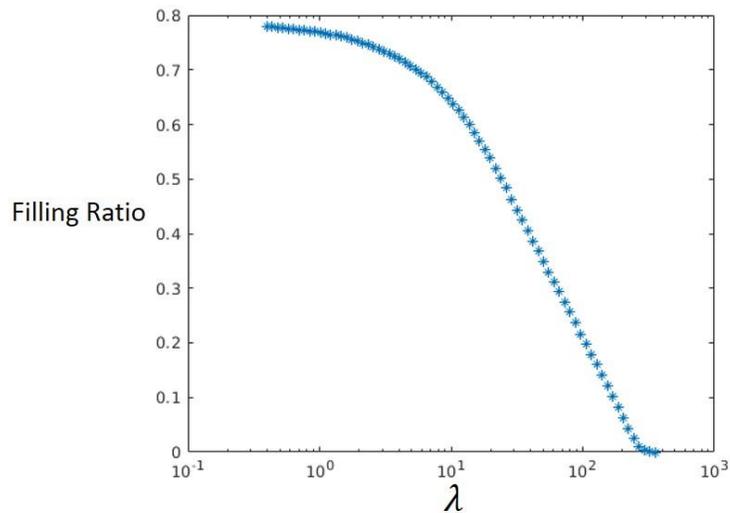

Figure 3: Trend of sparsity of the estimated filters against $\lambda$ values.

To verify prediction performances of the estimated $\widehat{w}$, we reconstructed the full intensity image with all (64 by 64) CBED data associated with the full graphene HAADF image (Figure 1a) containing a 4-fold silicon dopant. Fig 4 shows the reconstructed intensity images (first row) with the estimated filters (second row) for multiple $\lambda$ values. As $\lambda$ decreases, the sparsity of estimated filters decreases and more information inside CBED images are utilized. Smaller $\lambda$ values yield higher resolution meanwhile there is a larger discrepancy between the training area and test area of the reconstructed image. As $\lambda$ decreases, the training area of the reconstructed image could achieve the artificially high resolution, but the test area remains at a lower resolution. We also note that the intensity of the 4-fold silicon dopant (brighter spot) in the reconstructed image is still higher.

We further validated the performances of the same set of estimated filters in Figure 4 on another CBED dataset, collected under the same instrument settings but corresponding to a graphene HAADF image with a 3-fold silicon dopant. Fig 5 shows the reconstructed intensity images on this graphene with a 3-fold silicon dopant CBED dataset. Figure 6a is the original graphene HAADF image containing a 3-fold silicon dopant. It is clear from visual inspection of Fig. 5 that the C dumbbells are clearly resolved for lower $\lambda$ values, achieving a stable resolution level lower than that of the training area in Figure 1b. For higher $\lambda$ values the dumbbells become less distinct. The brighter spot in the reconstructed images preserves the 3-fold silicon dopant. Figure 6b compares line scans of reconstructed images and the original HAADF image. We have taken a line trace, 3 pixels wide, through adjacent dumbbells in Fig. 5 as illustrated by the blue boxes. In order to compare line traces of different intensity ranges, we have subtracted the minimum value from each image and normalized the maximum value of each line trace to one. These are compared to a similarly adjusted HAADF line trace in Fig. 6b. It can be seen that the reconstructions have comparable contrast to the simultaneously acquired HAADF image, but the higher $\lambda$ values lead to a systematic non-symmetry in the dumbbells. It needs to be noted that the limitations of the camera used here required large dwell times and so effects like sample drift will affect the images acquired here. Recent advances in direct electron detection cameras allow much faster dwell times and promise the possibility of resolutions approaching those seen for specialist ADF imaging in the literature [29], [30].

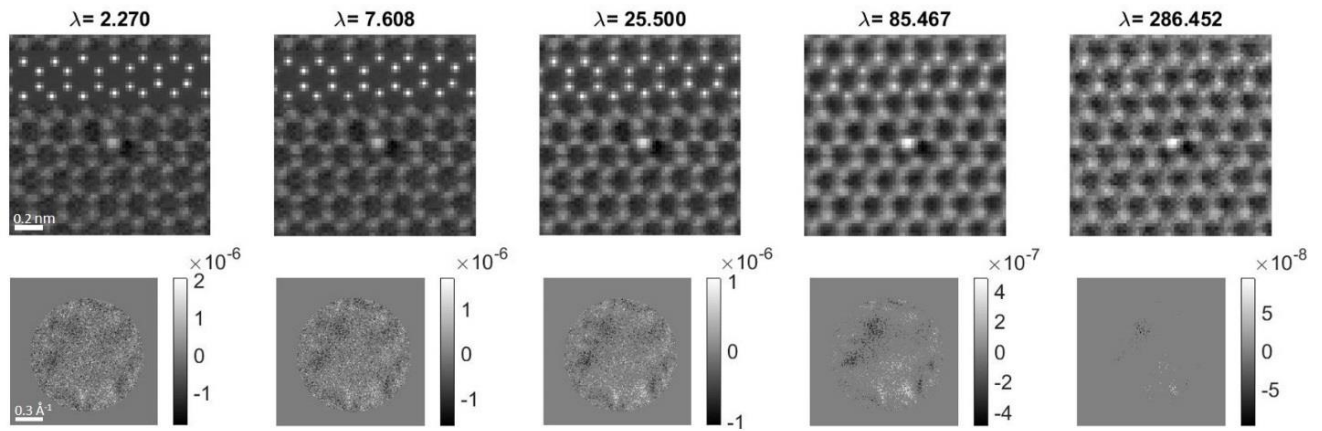

Figure 4: Evolution of reconstructed intensity images (top row) and the corresponding estimated filters (bottom row) along the increasing sequence (left to right) of $\lambda$ values.

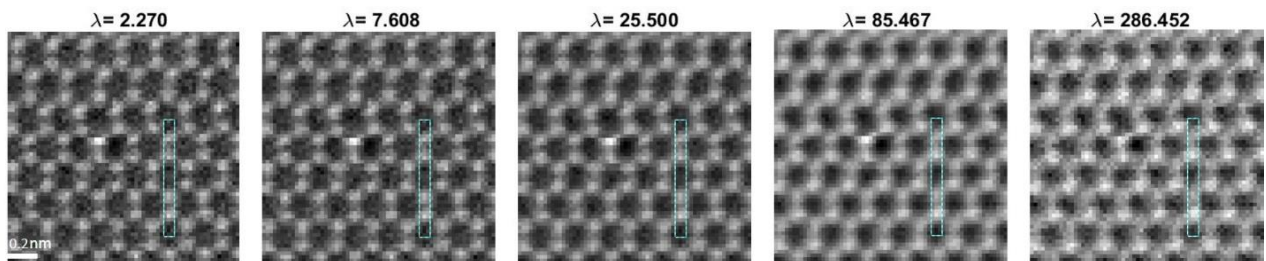

Figure 5: Evolution of reconstructed intensity images from CBED datasets associated with HAADF image of graphene with a 3-fold silicon dopant, by the same set of estimated filters as shown in the bottom row of Figure 4.

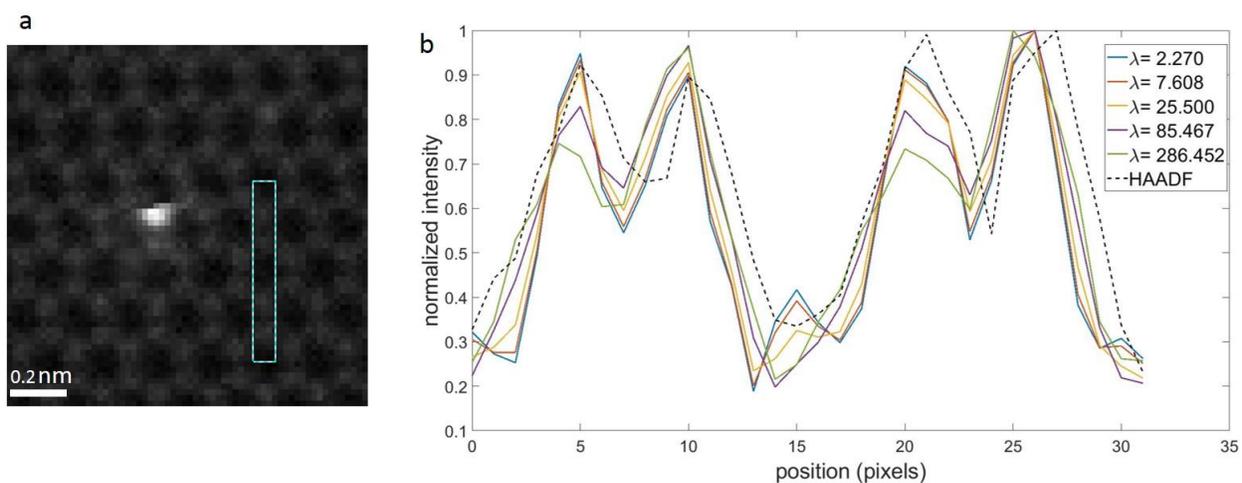

Figure 6: (a) Graphene HAADF image showing a 3-fold silicon dopant. (b) Averaged line scan profiles taken inside boxes in Figure 5 and in (a).

We then investigated how these techniques would apply for segmented detectors. We use the same training intensity image as in Figure 1b. For the associated training CBED data, we integrated over pixel ranges corresponding to each segment. We considered 8-, 16-, 32-, 64- and 128 segment detectors. For testing, we used the graphene with a 3-fold silicon dopant CBED dataset. Figure 7 shows results based on a small $\lambda = 0.035$. The top row of Figure 7 shows the reconstructed intensity images and the second row contains estimated filters. Reconstruction quality tends to improve as the total number of segments increases. Segmented detectors with more than 32 segments appear to have similar reconstruction performances compared to pixelated detectors. These findings are consistent with the previous analysis in Ref [16] where the optimal detector configurations were analyzed.

Next, we would like to find the optimal filter $w$ utilizing only the information outside the bright field disk of the CBED (we manually set the intensity inside the bright field disk to be 0). The training and testing framework is the same as above. The top row of Figure 8 shows the reconstructed intensity images with all CBED data associated with the 4-fold silicon dopant case. Correspondingly, the middle row of Figure 8 contains the estimated filters. The bottom row shows reconstructions from CBED data associated with the 3-fold silicon dopant case. Under this setup, we still can get good training results, but we failed to get any acceptable verification results on the testing CBED datasets. This failure of validation may imply that, for the CBED datasets considered here, the area outside bright field disk does not record enough information for STEM imaging.

In Figure 9, we used a synthetic intensity image with random atom positions (Figure 2b) for training. Still we can get perfect training results, but for validation performances, we could not get any meaningful reconstruction images. There is no clear pattern in the estimated filters.

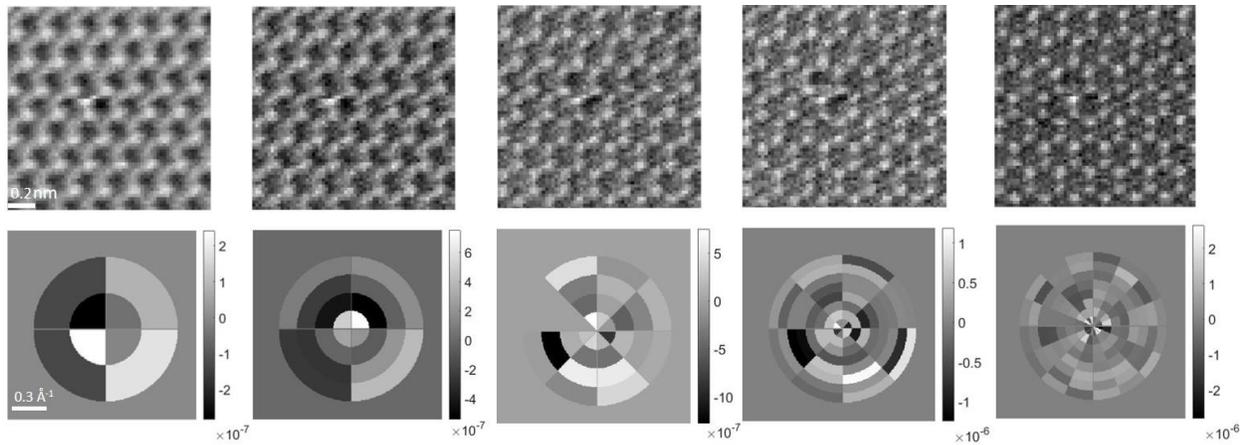

Figure 7: Reconstructions and estimated filters for segmented detector settings.

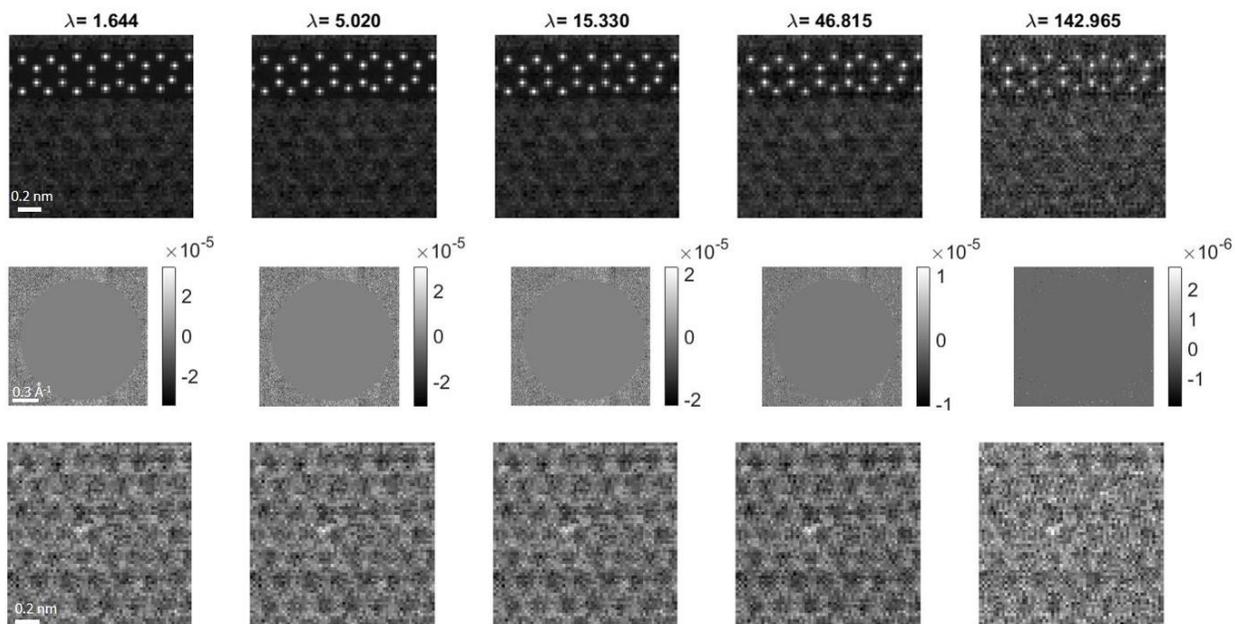

Figure 8: Reconstruction results via only using information outside bright field disk. Top row: reconstructed intensity images from CBED datasets associated with the 4-fold silicon dopant case. Middle row: estimated filters. Bottom row: reconstructed intensity images from CBED datasets associated with the 3-fold silicon dopant case.

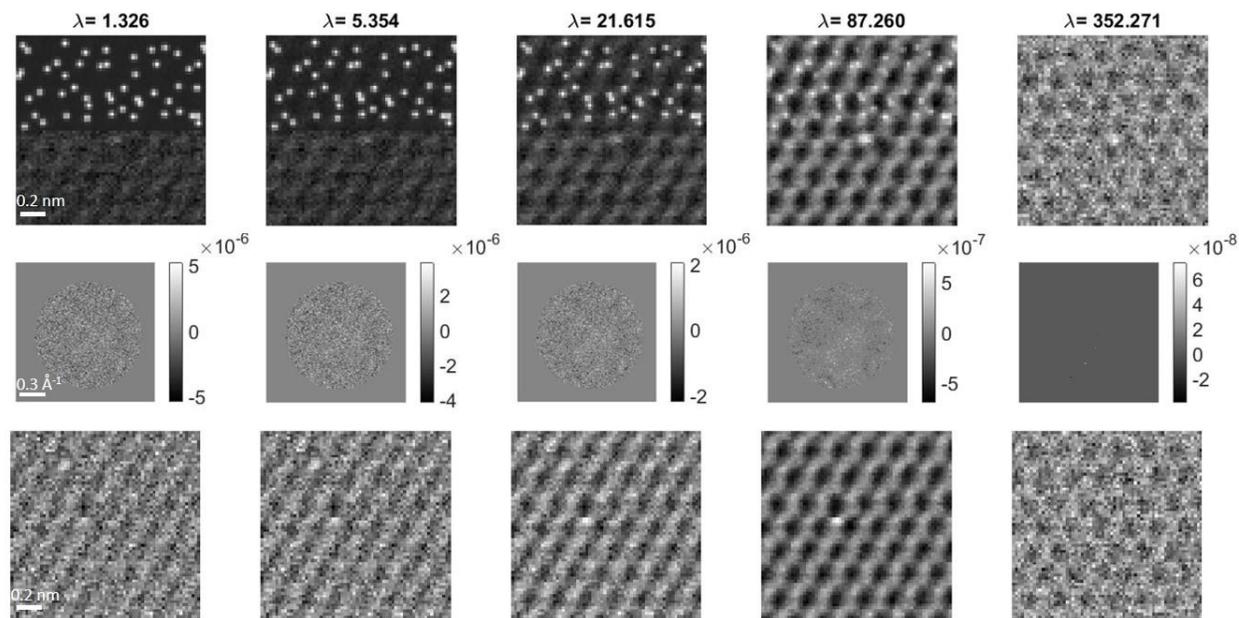

Figure 9: Results via training with intensity image containing random atom sites (as shown in Figure 2b). Top row: reconstructed intensity images from CBED datasets associated with the 4-fold silicon dopant case. Middle row: estimated filters. Bottom row: reconstructed intensity images from CBED datasets associated with the 3-fold silicon dopant case.

# Discussion

Since the first proposals by Rose and Dekkers and De Lang a half century ago, several groups have been making significant progress on finding optimal configurations of electron detectors for atomic structure retrieval. In this communication, we demonstrated that the wealth of information contained in CBED (even in only part of CBED pattern) could numerically yield almost arbitrary real-space structures that have meaningless interpretations. Thus attention should be paid to prediction and validation performances. We explored statistical model selection algorithms for finding robust filters on CBED datasets to reconstruct real-space imaging. We further showed that, by careful choice of model selection and validation, we could get robust filters on the CBED patterns, yielding comparable imaging as conventional HAADF imaging technique. Our analysis framework naturally unites the pixelated and segmented detector settings. Future work could involve exploiting advanced regularization terms for the inversion model, such as structural sparsity penalizations [31], [32], instead of the flattened L1+L2 norm penalty used in this paper.

Incorporating prior knowledge and physical constraints of 4D-STEM datasets into the inversion model via unsupervised learning and exploratory data analysis [33] could also possibly improve the robustness of estimated filters. Having a second look at the overfitted results in Figure 2, we note that, for training images Figure 2a and Figure 2c, the predictions (lower half of Figure 2i and Figure 2k) somewhat display the periodic structure, meanwhile the predictions totally fail for the non-periodic structures in the training images Figure 2b and Figure 2d. For the training image Figure 2b, Figure 9 further shows that adjusting the sparsity regularization did not help to get robust and interpretable predictions. We end this communication with three more examples obeying the periodicity constraint: 1. We used the complement of the training image in Figure 1b to mimic the virtual bright field image; 2. We decomposed the training image in Figure 1b to separately contain atom sites on graphene sub-lattices. Figure 10 shows the corresponding reconstruction results at small $\lambda$ values, where predictions preserve the periodicity, successfully recovering the bright-field image and either of the sub-lattices. We may conjecture that, for the graphene CBED datasets studied here, periodicity is a useful constraint for robust reconstruction. In other words, statistical model selection may provide measures and compatibility intervals for justifying physical hypothesis imposed on 4D-STEM virtual imaging.

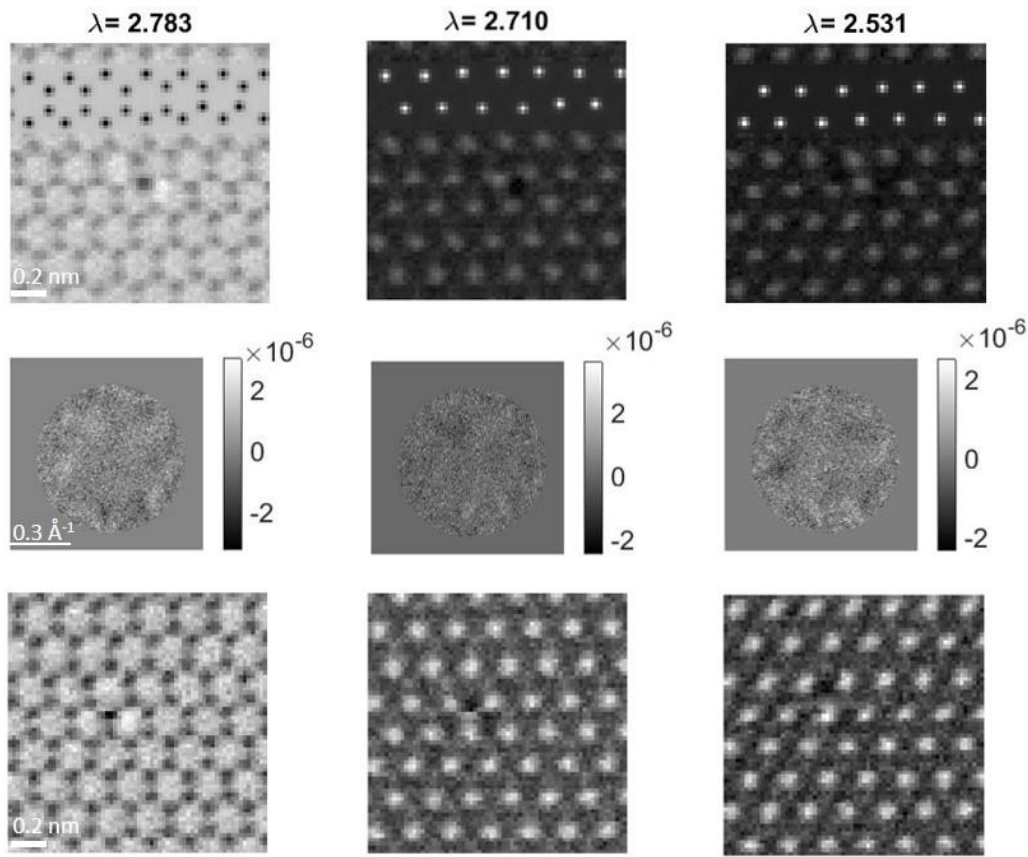

Figure 10: Reconstruction results via training images obeying periodicity structure: complement of Figure 1b (left column) and sub-lattices of Figure 1b (middle and right column). Top row: reconstructed intensity images from CBED datasets associated with the 4-fold silicon dopant case. Middle row: estimated filters. Bottom row: reconstructed intensity images from CBED datasets associated with the 3-fold silicon dopant case.

## Acknowledgements

This work was supported by the Laboratory Directed Research and Development program of Oak Ridge National Laboratory, managed by UT-Battelle, LLC for the U.S. Department of Energy (X.L., O.D., S.J.), Oak Ridge National Laboratory's Center for Nanophase Materials Sciences (CNMS), a U.S. Department of Energy Office of Science user facility and the U.S. Department of Energy, Office of Science, Basic Energy Sciences, Division of Materials Science and Engineering (M.P. O., A.R.L., S.V.K.).

.